# SIMPLE EXPERIMENT CONFIRMING THE NEGATIVE TEMPERATURE DEPENDENCE OF GRAVITY FORCE

Alexander L. Dmitriev

*St-Petersburg National Research University of Information Technologies,
Mechanics and Optics, 49 Kronverksky Prospect, St-Petersburg, 197101, Russia*

Results of weighing of the tight vessel containing a thermo-isolated copper sample heated by a tungstic spiral are submitted. The increase of temperature of a sample with masse of 28 g for about $10^0 C$ causes a reduction of its apparent weight for 0.7 mg. The basic sources of measurement errors are briefly considered, the expediency of researches of temperature dependence of gravity is recognized.

## Introduction

The question on influence of temperature of bodies on the force of their gravitational interaction was already raised from the times when the law of gravitation was formulated. The first exact experiments in this area were carried out at the end of XIXth - the beginning of XXth century with the purpose of checking the consequences of various electromagnetic theories of gravitation (Mi, Weber, Morozov) according to which the force of a gravitational attraction of bodies is increased with the growth of their absolute temperature [1-3]. That period of experimental researches was completed in 1923 by the publication of Shaw and Davy's work who concluded that the temperature dependence of gravitation forces does not exceed relative value of $2 \cdot 10^{-6} K^{-1}$ and can be is equal to zero [4]. Actually, as shown in [5], those authors confidently registered the negative temperature dependence of gravitation force. Nevertheless, probably in the view of growth of popularity of the general theory of relativity (GR) by Einstein, Shaw and Davy have not dared to insist on their results. According to GR doctrines, the positive relative temperature dependence of gravitation force has the order of $10^{-15} K^{-1}$ which means that it can not practically be observed [6].

It is necessary to note that non-relativistic, including ether models of gravitation, as well as separate (sometimes exotic) hypotheses of the nature of gravitation, do not exclude rather "strong" temperature dependence of the force of gravitation [7,8]. Here it is pertinent to note the experiment by A.P. Shchegolev made in 1983 who, following the idea of the "thermodynamic" nature of gravitation, had observed the temperature reduction of weight of the massive steel sphere which was heated up with a beam of a powerful laser [9]; unfortunately, the accuracy of those experiments was rather insignificant.

In the 90s, in the process of researches into the influence of the accelerated movement of a test body on the results of its exact weighing and on the basis of individual analogies of the gravitational and electromagnetic phenomena, the author and his colleagues had executed measurements of apparent weights of samples of non-magnetic metal rods exited by ultrasound [10]. The longitudinal ultrasonic waves in rods, created by the piezoelectric converter, are accompanied by significant accelerations of the rod material microparticles which fact was used as a basis of measurements idea. During the experiments it was found that the results of measurements of rod weights were significantly influenced by the increase of their temperature, owing to both absorption of ultrasound and heat transfer from the piezo-converter. The frequency spectrum of temperature fluctuations of solid body particles lies in the field of hypersound frequencies and essentially exceeds the frequency of ultrasound, therefore such results are natural. The measurements of temperature dependence of weight (apparent mass) of metal samples executed by the specified technique had shown a rather strong negative temperature dependence of their weight with relative value $\gamma$ in the range from $4.6 \cdot 10^{-6} K^{-1}$ for a lead sample up to $11.6 \cdot 10^{-6} K^{-1}$ for duralumin one. The elementary phenomenological theory of temperature dependence of samples weight [10,11] had given a satisfactorily explanation of temperature dependence factor $\gamma$ on density $\rho$ and elastic properties (speed $V$ of longitudinal ultrasonic waves) of material:

$$\gamma \propto -\frac{V}{\sqrt{\rho}} \quad ; \tag{1}$$

the same model allowed to prove the orientation dependence of weight of some crystals [12].

Though the results of measurements of negative temperature dependence of gravitation are in obvious disagreement with conclusions of GR, they do not contradict the earlier executed experiments or are challenged by anybody. On the contrary, in 2010 the work of Chinese physicists (Liangzao Fan, Jinsong Feng, Wuquing Liu) investigating the temperature dependence of weight of metal samples (including samples from copper, gold and silver) was published in which the sign and value of temperature factor $\gamma$ for a copper sample closely corresponded to our data [13]. Recently conducted measurements of temperature dependence of weight of the pile of piezoelectric converters excited on resonant frequency had also appreciably shown the reduction of piezo-ceramics weight with the growth of its temperature [14].

## Experiment

In the described experiment a tight vessel was weighed inside which there was a thermo-isolated copper sample heated by a tungstic spiral through which the electric current was passed. The design and appearance of the container are shown in Fig. 1 and Fig. 2. Diameter of the vessel is 63 mm, height - 87.5 mm, mass - 127.4 g.

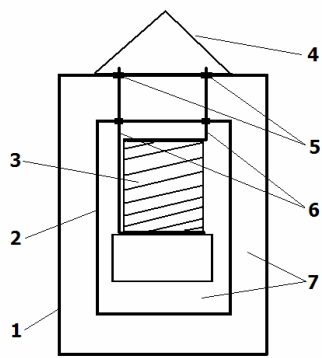

Fig. 1. The design of the container. 1 - external, sealed tin vessel; 2 - an envelope of the internal case (foil); 3 - copper cartridge wrapped up by mica and a tungstic spiral; 4 - a suspension bracket; 5 - «cold welding» (polymeric glue); 6 - copper conductors; 7 – thermo-isolation envelopes (foam plastic).

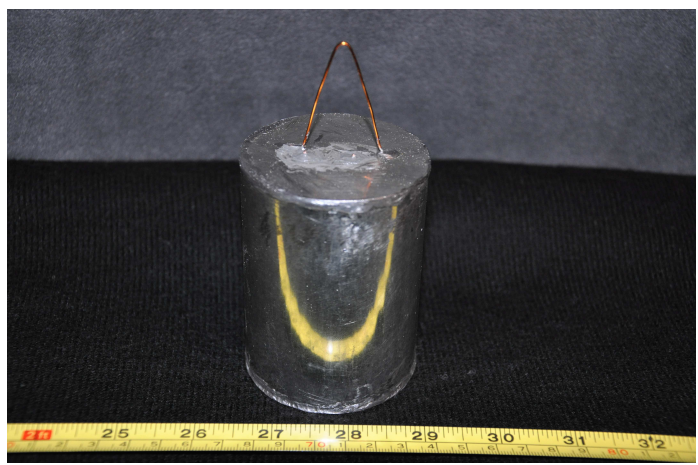

Fig. 2. Appearance of the container.

Weighing was made by the elongation method (with readout of extreme angular positions of oscillating scales beam) on laboratory scales of ADL-200 model. Procedure of weighing consisted of three stages. At the first stage the container with a cold sample was weighed for 5-6 minutes. At the second stage of total duration equal to 2.5 minutes the electrodes of the container heater were connected for 35 seconds to a power supply source; the current in the heater circuit is 0.9A, resistance of the winding - 6.5 Ohm. At the third stage the continuous readout of current values of container weight was made for 8-9 minute. All mentioned manipulations were repeatedly and carefully fulfilled, the resulting error of measurement of weight of the container did not exceed 50 mcg.

During measurements the temperature of walls and bottom of the container was equal to $24.9 \pm 0.1^0 C$. The temperature of the "hottest" central area of the container cover from the moment of the heater switching ON has grown with speed of less than $0.1^0/\min$ ; so, in the first 2 minutes after switching OFF of the heater power supply, the temperature of the container cover had increased by no more than $0.15^0 K$ . Under the specified conditions the influence of the air convection flows caused by difference of temperatures on the surface of the container and air in a closed box of scales was practically insignificant. Absence of a leak (hermetic sealing) in the container was controlled by usual methods. Typical time dependence of container apparent weight is shown in Fig. 3.

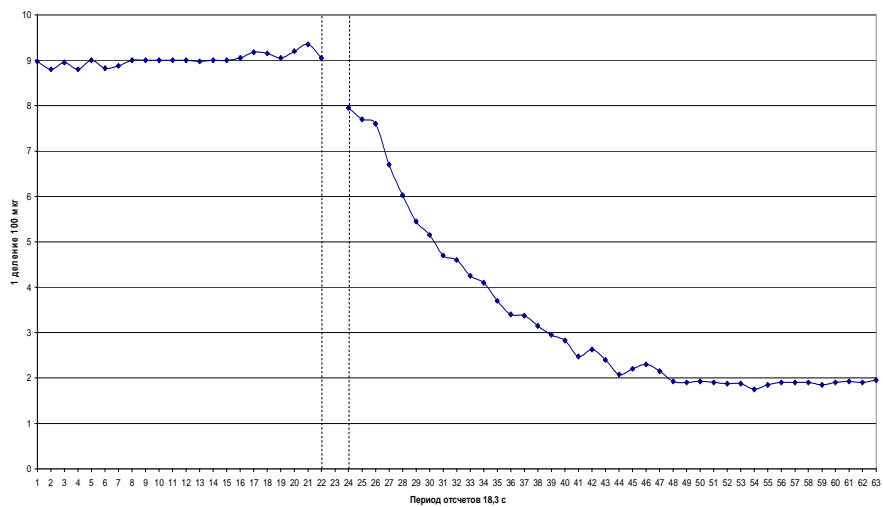

Fig. 3. Time dependence of changes of the container apparent weight. Shaped lines specify the moments of the beginning and the end of the second stage of weighing (the time scale of this stage is reduced). Vertical axis: 1 div 100 mcg; horizontal axis: period of sample 18.3 s.

During the first 2-3 minutes after switching OFF of the heater the reduction of the container weight is maximal, reaching 200 mcg, then during approximately 3 minutes a monotonous reduction of weight down to $\Delta m \approx 700 mcg$ is observed.

### Discussion

The calculated amount of heat supplied by the electric heater is approximately equal to 184 J. A part of this heat is dissipated in conductors and heat insulator, but significant, in amount of about 100J, part $\Delta Q$ is transferred to copper sample 3 (Fig. 1) which has mass $m \approx 28g$ . The respective change $\Delta T$ of average temperature of uniformly heated sample ($\Delta T = \Delta Q/mc$ , where $c = 3.9 \cdot 10^2 J/kg \cdot K$ is a specific thermal capacity of copper) $\Delta T \approx 9^0$ . The relative temperature change of sample weight,

$$\gamma = \frac{\Delta m}{m\Delta T} \quad , \qquad (2)$$

is equal to $\gamma = -2.8 \cdot 10^{-6} K^{-1}$. This value differs by more than 2 times from the result received in ultrasonic heating of a copper sample; nevertheless, the sign and the order of value $\gamma$ correspond to the previous measurements [10-14]; comparing the specified results it is necessary to consider the difference of physical conditions of heating the sample by ultrasound and in the process of heat transfer.

The general character of apparent weight reduction in Fig. 3 is explained by the process of distribution of heat in a copper sample of complex configuration (the heated part of the copper hollow cartridge makes approximately half of its length; the diameters of heated up and no-load parts of the cartridge differ) and, on the whole, correspond to similar dependence in measurements of weight of a brass core in Dewar's vessel [10]. In both cases the monotonous time dependence of the measured weights is explained by slow distribution of thermal wave in the samples being weighed.

As shown by the special measurements, the temperature change of the top part of the container during the experiment did not exceed $0.5^0$, and the temperature of walls remained constant with accuracy of $0.1^0$. Under these conditions, on the basis of Glaser's theory [15], the changes of apparent weight of the container, caused by air convection, and the change of buoyancy of the sample do not exceed $50 mcg$. So, with a difference of air temperature $\Delta t$ and temperatures of the cylinder surface of the area $A = 173 cm^2$ and the diameter $d = 6.3 cm$, the change $\Delta m$ of apparent weight of a cylindrical sample is equal to

$$\Delta m = -9.2 \cdot 10^{-7} A d^{1/4} \Delta t^{3/4} \qquad . \qquad (3)$$

With $\Delta t = 0.2^0$ from equation (3) which is obviously overestimated for conditions of the described experiment it follows that $\Delta m \approx 75 mcg$ which is approximately by an order less than the total value of the observed weight change.

It can be shown that the change of buoyancy caused by temperature change of the container volume owing to temperature change of the container case even by $1^0$, causes a measurement error of its apparent weight of less than 10 mcg.

On the whole, the results of the executed experiment will be coordinated with the dates [10-13] received earlier, confirming the fact of rather strong negative temperature dependence of force of the gravitation acting on the heated test body.

*** 

As a note, we shall mark the following. If during the further experiments it will be shown that the specified temperature dependence has the universal character there probably will be necessary to update some conclusions of the well-known theories and models of gravitation.

In particular, the negative temperature dependence of gravitation force specifies that in the course of (astrophysical) gravitational collapse accompanying by an increase of temperature of collapse-mass, the realization of the so-called condition of "singular point" is impossible. Hence, the popular hypothesis of "black holes" can seem to be rather doubtful.

## Conclusions

1. An increase of temperature of thermo-isolated copper sample with mass of 28 g by value of about $10^0 C$ is accompanied by reduction of its apparent weight by more than 0.7 mg.

2. The observed, rather "strong" negative temperature dependence of body physical weight with relative value by the order of $10^{-6} K^{-1}$ does not contradict the known experiments on exact weighing the heated test body.

3. The experimental researches of temperature dependence of weight of bodies of the various structures, conducted in a wide range of temperatures, will promote the development of concepts of physics of gravitational interaction.